\documentclass[aps,prl,twocolumn,showpacs,superscriptaddress]{revtex4-1}
\usepackage[utf8]{inputenc}
\usepackage{amsmath}
\usepackage{amsthm}
\usepackage{amssymb}
\usepackage{lmodern}
\usepackage{multirow}
\usepackage{booktabs}
\usepackage{graphicx}
\usepackage{placeins}
\usepackage[capitalise]{cleveref}
\usepackage{xcolor}
\usepackage{lineno}

\begin{document}
\title{Transferability of Deep Learning Models in Searches for New Physics at Colliders}
\author{M. Crispim Rom\~{a}o}
\email{mcromao@lip.pt}
\affiliation{LIP, Av. Professor Gama Pinto 2, 1649-003 Lisboa, Portugal}
\author{N. F. Castro}
\email{nuno.castro@cern.ch}
\affiliation{LIP, Av. Professor Gama Pinto 2, 1649-003 Lisboa, Portugal}
\affiliation{Departamento de F\'{i}sica, Escola de Ci\^{e}ncias, Universidade do Minho, 4710-057 Braga, Portugal}
\author{R. Pedro}
\email{rute@lip.pt}
\affiliation{LIP, Av. Professor Gama Pinto 2, 1649-003 Lisboa, Portugal}
\author{T. Vale}
\email{tiago.vale@cern.ch}
\affiliation{LIP, Av. Professor Gama Pinto 2, 1649-003 Lisboa, Portugal}
\affiliation{Departamento de F\'{i}sica, Escola de Ci\^{e}ncias, Universidade do Minho, 4710-057 Braga, Portugal}

\date{\today}

\begin{abstract}
In this work we assess the transferability of deep learning models to detect beyond the standard model signals. For this we trained Deep Neural Networks on three different signal models: $tZ$ production via a flavour changing neutral current, pair-production of vector-like $T$-quarks via standard model gluon fusion and via a heavy gluon decay in a grid of 3 mass points: 1, 1.2 and 1.4~TeV. These networks were trained with $t\bar{t}$, $Z$+jets and dibosons as the main backgrounds. Limits were derived for each signal benchmark using the inference of networks trained on each signal independently, so that we can quantify the degradation of their discriminative power across different signal processes. We determine that the limits are compatible within uncertainties for all networks trained on signals with vector-like $T$-quarks, whether they are produced via heavy gluon decay or standard model gluon fusion. The network trained on flavour changing neutral current signal, while struggling the most on the other signals, still produce reasonable limits. These results indicate that deep learning models are capable of providing sensitivity in the search for new physics even if it manifests itself in models not assumed during training. 
\end{abstract}

\maketitle

\section{Introduction}

Although machine learning has a long history in High Energy Physics (HEP), we have recently witnessed a surge in interest in new methods and algorithms emerging from deep learning~\cite{Guest:2018yhq}. Deep learning models differ from those of traditional machine learning as they are composed of a stack of layers with non-linear functions that have the capacity to learn hierarchical features from the inputs~\cite{Goodfellow-et-al-2016}. Indeed, it has been shown that deep learning models for computer version trained on a certain task can be adapted to a different, albeit similar, task~\cite{yosinski2014transferable} as the layers closer to the inputs learn low-level features that progressively become higher-level as they are transformed by the subsequent layers. In computer vision this manifests as the first layers learn about localised pixel variations, the following learn about textures and patterns, and finally the last encode high-level features such as \emph{dog} or \emph{cat}. This has led to reuse deep learning models trained on a specific task, say to discriminate between dogs and cats, to perform a different one, say to discriminate between cars and trucks, by keeping the lower layers and train or fine-tune the layers responsible for the high-level features. A natural question then arises if the same transferability happens on deep learning models used in HEP. More specifically, how transferable is a deep learning model trained on reconstructed physical observables in the discriminative task of separating signal from background when different signals are considered? Since the task is the same we will not retrain or fine-tune the last layers. Instead, we will assess how a trained model can perform the same task given a different signal sample.

The goal of this work is to study how deep learning models trained on a specific signal are transferable to new signals unseen during training. As such, a few benchmark signals were considered, having in common the presence of $tZ+X$ final states. Since the target topology determines which Standard Model (SM) backgrounds have to be considered, we focused on events with at least a pair of leptons, arising from the decay of the $Z$-boson or top-quark, as well as with jets originating from $b$-quarks, from the decay of the top-quark. Three classes of well motivated signals were considered in this context: (a) $tZ$ production via effective field theory operators inducing a flavour changing neutral current (FCNC) coupling $utZ$; (b) pair production of vector-like $T$-quarks (VLT) via a SM gluon and (c) pair production of vector-like $T$-quarks via a beyond-SM heavy gluon. By testing the ability of the model to recognize these new signal events, we probe an intermediate step of generic detection, in the midway between fully signal-tailored classifiers and anomaly detectors.

The remainder of the paper is organised as follows. We start by briefly describing and motivating the BSM phenomenological grounds covering the new physics signals explored in the paper. The following section presents the data used in the analysis, from signal and background simulation to event selection and data preparation. Afterwards we present the Deep Neural Network (DNN) model and the procedure used for its architecture optimization. Results are presented and discussed in the section that follows, with conclusions being drawn at the end.

\section{Beyond Standard Model signals}

FCNC interactions in single top-quark production can appear in dimension-four or dimension-five operators of an effective SM extension. The search for FCNC processes involving top-quarks and $Z$-bosons at the Large Hadron Collider (LHC) has been performed considering both $t\bar t$ production with $t\to qZ$ decays (with $q$ being an $u$- or $c$-quark)~\cite{Aaboud:2018nyl,Chatrchyan:2013nwa} and $tZ$ production~\cite{Sirunyan:2017kkr}. In the current paper we focus on the $tZ$ production via a $utZ$ FCNC coupling~\cite{FCNCmodel1,FCNCmodel2} since it produces a final state belonging to the $tZ+X$ category we are targeting.

Both the ATLAS and CMS Collaborations have developed a comprehensive program to search for vector-like quarks at the LHC, targeting an extensive list of different final states. Among those, multi-lepton final states originated by the pair production of $T$-quarks, where at least one of them decays into $tZ$, were performed and stringent limits on the mass of the new quarks as a function of the branching ratio (BR) for the considered decays were obtained~\cite{Aaboud:2017qpr, Sirunyan:2018qau}. We focus on this $tZ$ final state, considering the pair production of $T$~quarks via a SM gluon and assuming that $T$ belongs to a weak isospin doublet, where the BR($T\to tZ$) is approximately $1/2$~\cite{VLQmodel,TopPartners}. In order to illustrate the effect on the signal kinematics, different $T$ masses were considered: $1.0$, $1.2$ and $1.4$~TeV.

One class of models predicting the existence of vector-like $T$-quarks are the composite Higgs Models~\cite{Kaplan:1983fs} with partial compositeness~\cite{Kaplan:1991dc}. In such models, new massive colour octets are naturally expected and thus the pair production of $T$-quarks can occur via this new heavy gluon, producing events with kinematic properties expected to be different from the $T\bar{T}$ production via a SM gluon~\cite{HGinterpretation, hg_proc}. In the current paper we consider as benchmark a heavy gluon with a mass of $3.0$~TeV and the same $T$ masses and weak isospin charge as for the standard production case.

\hbox{}

\section{Simulated samples}

We use simulated samples of proton-proton collision events generated with MADGRAPH5\_MCATNLO 2.6.5~\cite{madgraph} at leading order with a centre-of-mass energy of 13~TeV. The parton showering and hadronisation was performed with Pythia 8.2~\cite{pythia}, using the CMS underlying event tune CUETP8M1~\cite{CUETP8M1} and the NNPDF 2.3~\cite{NNPDF2.3} parton distribution functions. The detector simulation employs the Delphes 3~\cite{delphes} multipurpose detector simulator with the default configuration, corresponding to the parameters of the CMS detector.

We target processes with final state composed of at least two leptons (\emph{i.e.} electrons or muons), at least one jet identified as originated by the fragmentation of a $b$-quark and large scalar sum of transverse momentum ($p_T$) of all reconstructed particles in the event ($H_T>500$~GeV)\footnote{The transverse plane is defined with respect to the proton colliding beams.}, motivated by a generic search for large mass resonances with intermediate products involving top-quarks and $W/Z$ vector bosons in the decay chain. Our main source of background is therefore composed of $Z$+jets, top pair ($t\bar{t}$) production and dibosons ($WW$, $WZ$ and $ZZ$). To obtain a robust statistical representation of the backgrounds across a significant region of the phase space, and specially in the high $H_T$ limit where most of the signal is expected to be, we generate each of the mentioned backgrounds in ranges of kinematic properties by applying an event filter at parton level according to:

\begin{itemize}
  \item The top/anti-top $p_T$ ($p_T^{top}$) for $t\bar{t}$: $p_T^{top}<100$~GeV, $p_T^{top}\in[100,250]$~GeV, $p_T^{top}>250$~GeV;
  \item The scalar sum of the $p_T$ of the hard-scatter outgoing particles for $Z$+jets: $H_T<250$~GeV, $H_T\in[250,500]$~GeV, $H_T>500$~GeV;
  \item $W/Z$ $p_T$ ($p_T^{W/Z}$) for dibosons: $p_T^{W/Z}<250$~GeV, $p_T^{W/Z}\in[250,500]$~GeV, $p_T^{W/Z}>500$~GeV.
\end{itemize}

Over 18~M events were simulated: 500~k per signal sample, 8~M for $Z+$jets, 3~M for $t\bar{t}$ and 1.5~M per diboson sample.
The background and signal samples are normalized to their expected yield after selection using each process generation cross-section at leading order, computed with MADGRAPH5, matching a target luminosity of 150~fb$^{-1}$. The full data set statistics is then split into three subsets according to the 0.3:0.2:0.5 proportion, to be used by the DNN for training, validation and test, respectively. 

\begin{figure}[h]
  \caption{\label{fig:ht_fatjet1pt}Distribution of the (top) $H_T$ and (bottom) leading large-radius jet $p_T$ for total background and each signal type: $tZ$ production by FCNC, and $T\bar{T}$ production via heavy guon or without heavy gluon for $m_T=\{1.0,1.2,1.4\}$~TeV. The distributions are normalised to the generation cross-section and to an integrated luminosity of 150~fb$^{-1}$.}
  \includegraphics[width=0.49\textwidth]{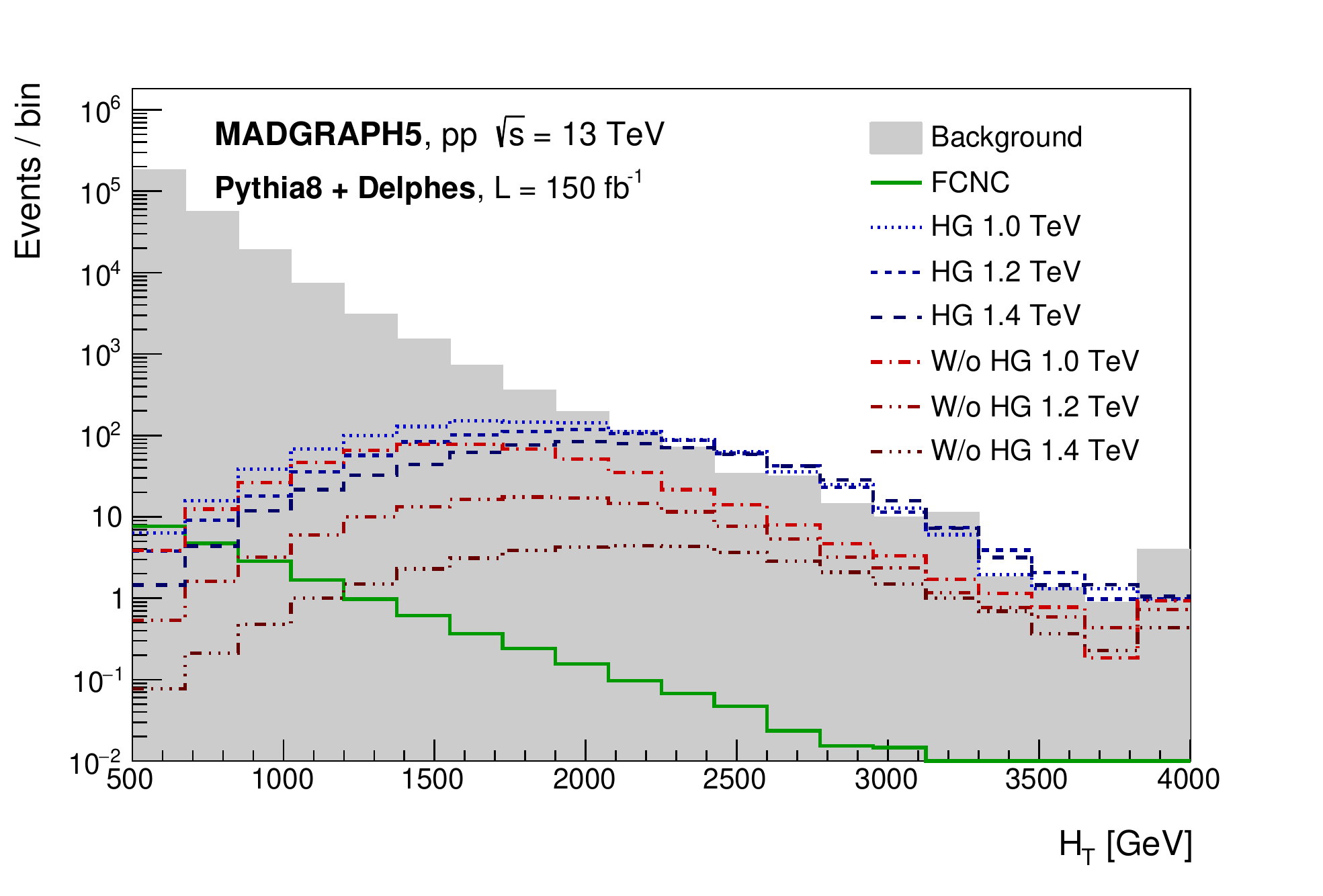}\\
  \includegraphics[width=0.49\textwidth]{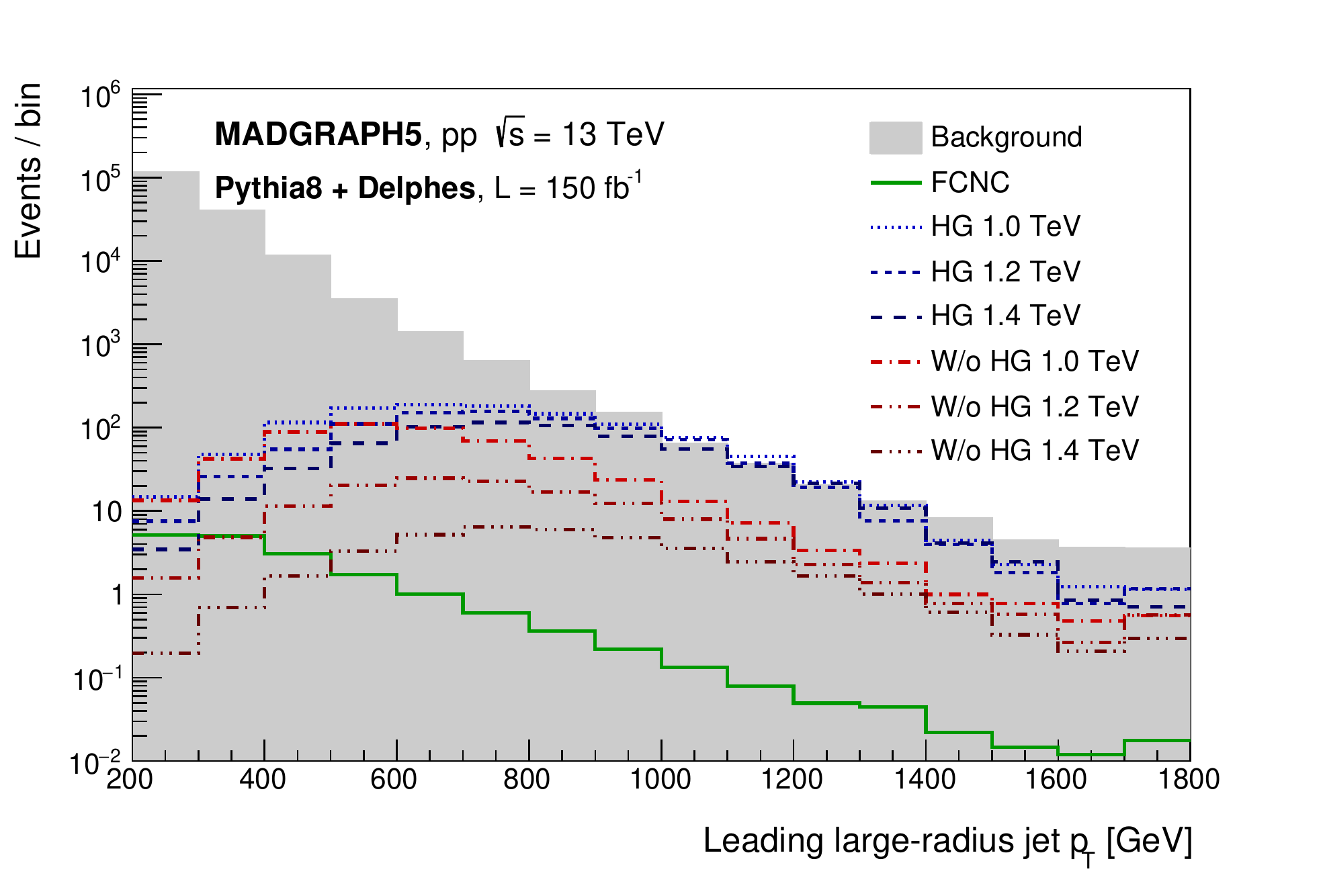}
\end{figure}

\cref{fig:ht_fatjet1pt} shows the distribution of $H_T$ and leading large-radius jet $p_T$ for the total simulated background and each signal. $T\bar{T}$ production via SM gluon exhibits significant kinematic dependency on the vector-like $T$ mass, with heavier $T$ processes producing harder final state objects. On the contrary, the kinematics of $T\bar{T}$ production by the heavy gluon is determined by the 3.0~TeV BSM gluon, much heavier than any of the $m_T$ cases, and is thus less sensitive to the $T$-quark mass. Nonetheless, there are substantial kinematic differences between the $T\bar{T}$ production via a heavy gluon when compared to the standard production case. While $T\bar{T}$ is characterised by large momenta final states, FCNC is on average closer to the SM background processes in terms of $H_T$ and momenta of the final state objects.

\section{Deep Neural Network Architecture and Training}

We train a DNN to distinguish each signal type from the set of backgrounds, using basic information constituted of the four-momenta of the reconstructed particles as provided by the Delphes simulation: 

\begin{itemize}
\item $(\eta,\phi,p_T,m)$ of the 5 leading jets and large-radius jets;
\item $(\eta,\phi,p_T)$ of the 2 leading electrons and muons;
\item Multiplicity of jets, large-radius jets, electrons and muons;
\item $(E_T,\phi)$ of the missing transverse energy ($MET$);
\end{itemize} 

\noindent where $\phi$ is the azimuthal angle defined in the transverse plane, $\eta$ is the pseudo-rapidity and $m$ the invariant mass. $E_T$ is the energy in the transverse plane. Jets are built from calorimeter energy clusters grouped using the jet finder algorithm anti-$k_t$ with radius parameter $R$=0.4 and $R$=1.0 for jets and in large-radius jets, respectively. Additional inputs are the $H_T$ and the $N-$subjetiness of the leading large-radius jet ($\tau_N$ with $N={1,2,..,5}$)~\cite{Marzani:2019hun}. 

DNNs are defined by a set of parameters, called hyperparameters, that specify their architecture. Although bigger networks (those with more consecutive layers and units per layer) provide greater representational and approximating power, they can also overfit to training data and therefore lose generalisation on new, unseen, data. Therefore, it is an important step in the DL workflow to assess the best configuration for the DNN for the task at hand. In order to find the best architecture, we implemented a Bayesian optimisation procedure using Scikit-Optimize~\cite{skopt}. The Deep Neural Networks themselves were implemented in Keras~\cite{Keras} on top of TensorFlow~\cite{TF}. Preprocessing and general data manipulation were performed using Numpy~\cite{Numpy}, Pandas~\cite{pandas}, and Scikit-Learn~\cite{scikit-learn}.

\begin{table}
  \caption{\label{tab:hyperparameters}Hyperparameters used by all DNNs.}
  \begin{ruledtabular}
    \begin{tabular}{ll}
      Hyperparameter           & Value        \\ \hline
      Hidden Layers            & 3            \\
      Units                    & 352          \\
      Unit Activation Function & Selu         \\
      Unit Weights Initialiser & LeCun Normal \\
      Dropout Rate             & 10\%         \\
      Initial Learning Rate    & $10^{-3}$    \\
      Optimizer                & Nadam        \\
      Maximum Epochs           & 1000
    \end{tabular}
  \end{ruledtabular}
\end{table}

At each iteration of the Bayesian optimisation loop, the performance of the network was assessed by the area under the receiver operating characteristics (ROC) curve on the valuation set. We compared the best architecture for each signal and found them to be very similar across the different signals. As a result, we fixed the architecture present in \cref{tab:hyperparameters} for all signals, and trained a network on each signal. Training was performed on batches of 2048 events and made use of Early Stopping after 15 epochs without improvement, while the learning rate was reduced by a factor of three after five epochs without improvement. Physical weights were used as training sample weights after being class-wise normalised, \emph{i.e.}:
\begin{equation}
  \sum_i^{N_b} \tilde w^b_i = \sum_i^{N_s} \tilde w^s_i
\end{equation}
where $\tilde w^b_i$ ($\tilde w^s_i$) and $N_b$ ($N_s$) are the training weights and number of events of the background (signal). We notice that this retains the relative weight ratios for different sub-samples within each class in the sample. The value of the sum is irrelevant, as it can be absorbed into the learning-rate, and we set it to 1.

\section{Results of Exclusion Limits}

With the trained networks, we proceeded to compute the predictions on all test samples, each comprised of the same background sample and a signal sample. The output is read as the probability of an event being of the signal in which the network was trained, \emph{i.e.}:
\begin{equation}
  \text{DNN}_S(X_i) = p(\text{S}| X_i) \ ,
\end{equation}
where DNN$_S$ is the Deep Neural Network trained on signal $S$ and $X_i$ the reconstructed variables of the event $i$.

\begin{figure}[htb]
  \caption{\label{fig:dnn_output}Distribution of the predictions for total background and each signal type, for the network trained on the $T\bar{T}$ signal produced via SM with $m_T=1.4$~TeV. The distributions are normalised to the generation cross-section and to an integrated luminosity of 150~fb$^{-1}$.}
  \includegraphics[width=0.49\textwidth]{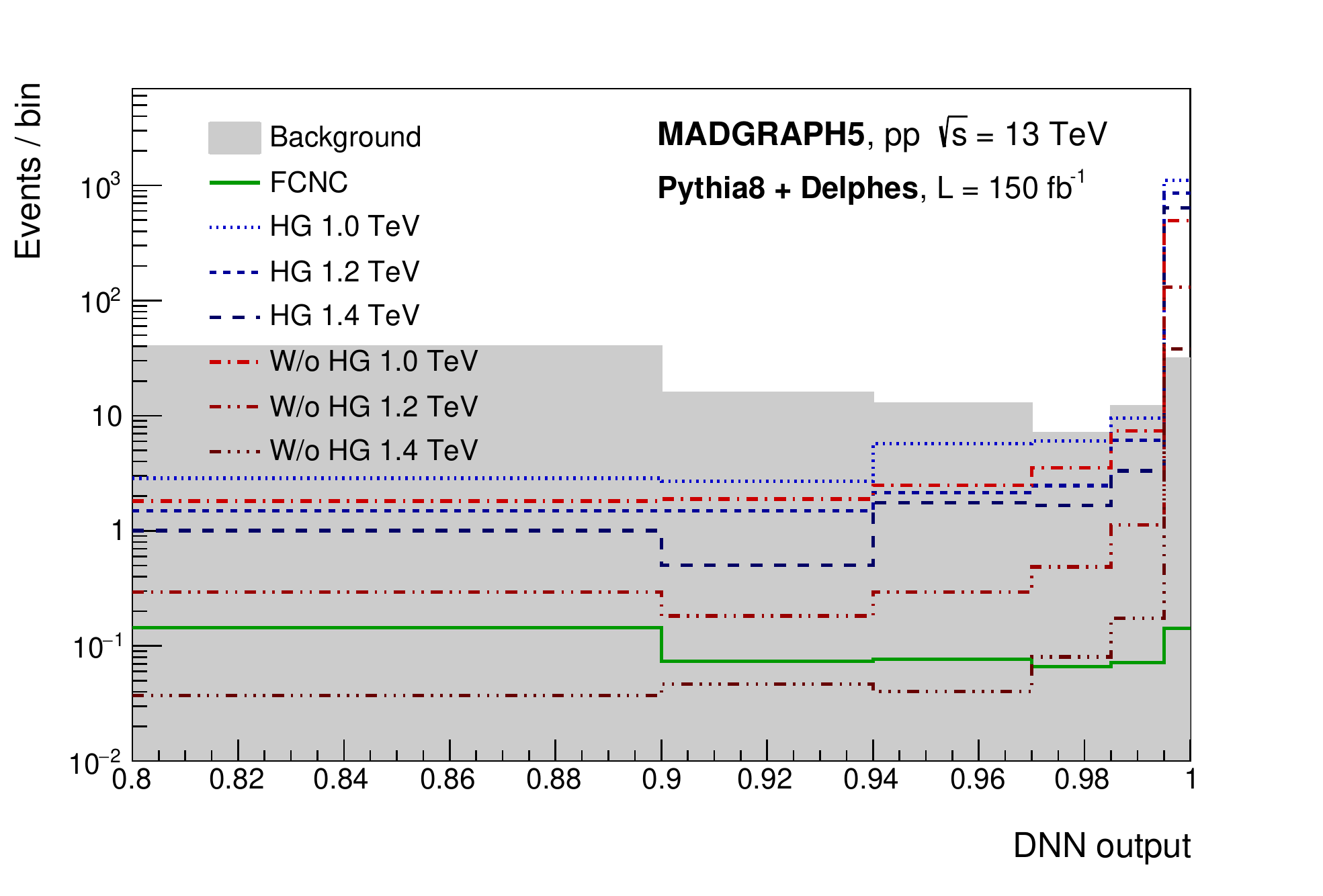}
\end{figure}

\begin{figure}[htb]
  \caption{\label{fig:roc_curves}Receiver Operating Characteristic (ROC) Curves for FCNC signal as predicted by all models (up), and for all signals as predicted by the model trained on $T\bar{T}$ produced via SM with $m_T=1.4$~TeV signal (down), where we notice that the curve for the FCNC signal is not shown with this level of zoom.}
  \includegraphics[width=0.45\textwidth]{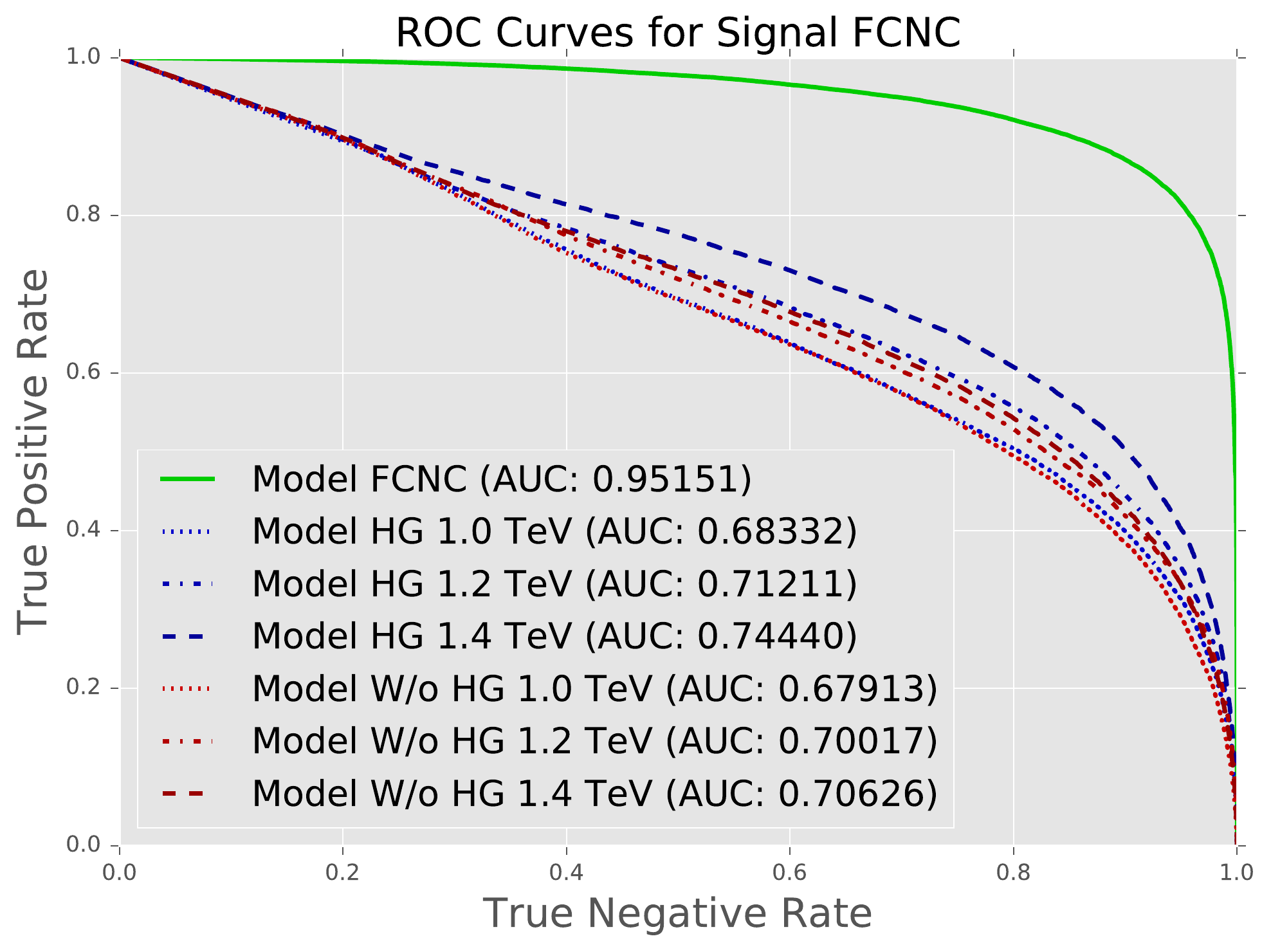}\\
  \includegraphics[width=0.45\textwidth]{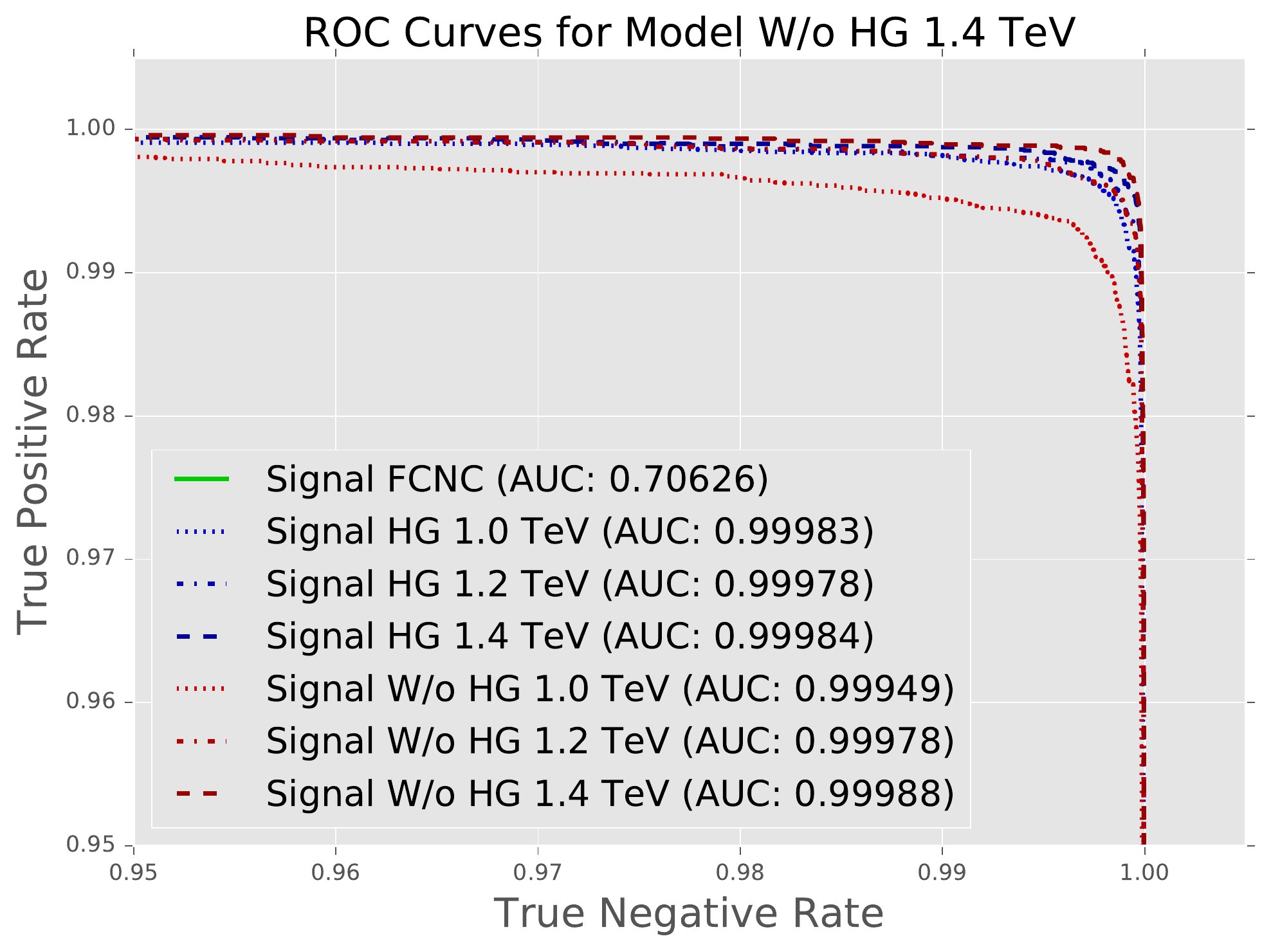}
  
\end{figure}

The prediction distributions of the network trained on the $T\bar{T}$ produced via SM with $m_T=1.4$~TeV for the background and each test signal is shown in  \cref{fig:dnn_output}. Although the network is unequivocally more performant in classifying the signal it was trained on, it is still able to identify other signal types as non-background. The quality with which this is done for each signal matches what is expected from the kinematic differences of the signals, presented in \cref{fig:ht_fatjet1pt}. In \cref{fig:roc_curves} we present Receiver Operating Characteristic (ROC) curves that show how the discriminating power degradates as we change the model used to predict signal, which is quite evident for FCNC signal. The bottom figure corresponds to the ROC curves of the predictions shown in~\cref{fig:dnn_output}.

\begin{figure}[htb]
  \caption{\label{fig:limits_heatmap}Heatmap of the central value of the limits on $\mu$ obtained for all combinations of training and testing signals, as presented in \cref{tab:limits}.}
  \includegraphics[width=0.45\textwidth]{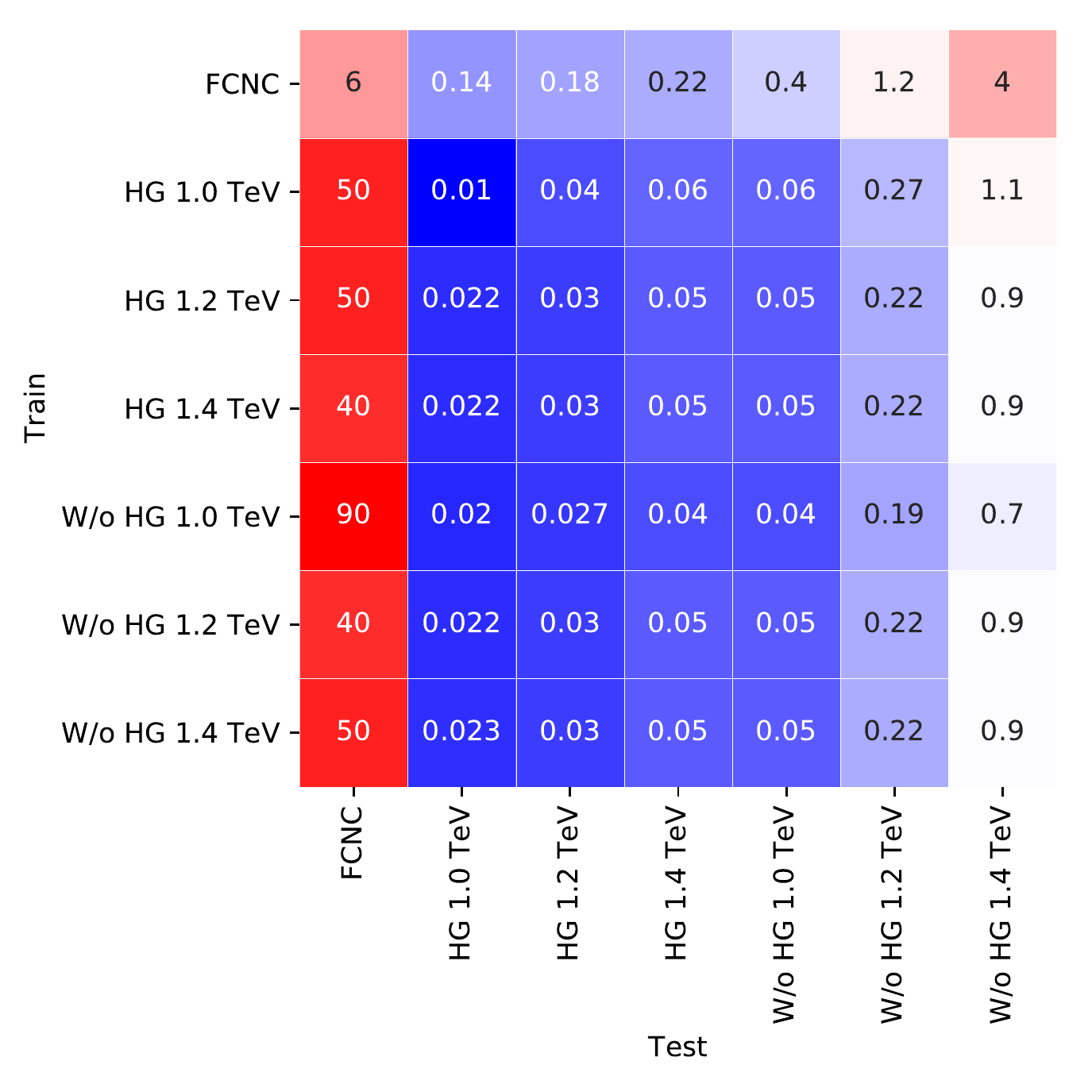}
\end{figure}

These distributions are used as the inputs of OpThyLic~\cite{opthylic}, a software to compute limits using the CL$_s$ method~\cite{Read:2002hq}. Poissonian statistical uncertainties on each bin of the distributions were included in the limit computation. From this we obtained the $95\%$ upper limit on the signal strength, defined as the signal cross-section normalized to the predicted cross-section, computed with MADGRAPH5, $\mu$. This procedure was done for each combination of train and test samples and the results are presented in \cref{tab:limits}. The central values presented in the table are used to produce the heatmap in \cref{fig:limits_heatmap}. The quoted uncertainties correspond to $1\sigma$ variations around the expected limit and incorporate the statistical uncertainty on the generated signal and background samples. We also present the value of the limit obtained for the fit to the $H_T$ distribution as a baseline to highlight how the discriminative power of the DNNs improves the sensitivity.

As expected, we observe that the signals with stricter limits are those with lower VLT masses for each case of either SM or Heavy Gluon production. A more interesting observation is that the limits seem to be reasonably insensitive to which VLT signal we used to train the discriminative DNN. 

The limits computed on the sample with FCNC signal show clear degradation as we use any network trained on VLT signals. This is easily understood as the FCNC does not produce new heavy states, and as such its kinematics are manifestly different from those produced by VLT signals, being instead very similar to other SM processes.

For comparison, \cref{tab:limits} also shows the limits for each signal type obtained by fitting the $H_T$ distribution as a simpler, but commonly used~\cite{Aaboud:2017qpr}, alternative to the use of DNN. We observe that for each signal, the limits obtained by employing a DNN are always better or compatible within the $1\sigma$ uncertainty than the ones obtained through the $H_T$ fit, regardless of the signal the model was trained on. This provides evidence that the usage of DNNs as discriminants can still provide impactful discrimination over signals not used during training.

\onecolumngrid

\begin{table}[b!]
  \caption{\label{tab:limits}Upper limits on signal strength, $\mu$, from the fit to the DNN output distribution for all combinations of train and test signals, and from the fit to the $H_T$ distribution.}
  \begin{ruledtabular}
    \begin{tabular}{ccclllllll}
      \multicolumn{1}{l}{}   & \multicolumn{1}{l}{}     & \multicolumn{1}{l}{} & \multicolumn{7}{c}{Test}                                                                                                                                                                                                                              \\
      \multicolumn{1}{l}{}   & \multicolumn{1}{l}{}     & \multicolumn{1}{l}{} & \multicolumn{1}{c}{\multirow{2}{*}{FCNC}} & \multicolumn{3}{c}{HG}            & \multicolumn{3}{c}{No HG}                                                                                                                                             \\
      \multicolumn{1}{l}{}   & \multicolumn{1}{l}{}     & \multicolumn{1}{l}{} & \multicolumn{1}{c}{}                      & \multicolumn{1}{c}{HG, $1.0$ TeV} & \multicolumn{1}{c}{HG, $1.2$ TeV} & \multicolumn{1}{c}{HG, $1.4$ TeV} & \multicolumn{1}{c}{$1.0$ TeV} & \multicolumn{1}{c}{$1.2$ TeV} & \multicolumn{1}{c}{$1.4$ TeV} \\
      \hline
      \multirow{7}{*}{Train} & \multicolumn{2}{c}{FCNC} & $6^{+2}_{-2}$        & $0.14^{+0.07}_{-0.04}$                    & $0.18^{+0.08}_{-0.06}$            & $0.22^{+0.10}_{-0.06}$            & $0.4^{+0.2}_{-0.1}$               & $1.2^{+0.5}_{-0.4}$           & $4^{+1}_{-2}$                                                 \\
                             & \multirow{3}{*}{HG}      & $1.0$ TeV            & $50^{+20}_{-20}$                          & $0.03^{+0.01}_{-0.01}$            & $0.04^{+0.02}_{-0.01}$            & $0.06^{+0.04}_{-0.02}$            & $0.06^{+0.03}_{-0.02}$        & $0.27^{+0.15}_{-0.09}$        & $1.1^{+0.6}_{-0.3}$           \\
                             &                          & $1.2$ TeV            & $50^{+20}_{-20}$                          & $0.022^{+0.011}_{-0.007}$         & $0.03^{+0.02}_{-0.01}$            & $0.05^{+0.03}_{-0.02}$            & $0.05^{+0.02}_{-0.02}$        & $0.22^{+0.11}_{-0.07}$        & $0.9^{+0.5}_{-0.3}$           \\
                             &                          & $1.4$ TeV            & $40^{+20}_{-10}$                          & $0.022^{+0.012}_{-0.007}$         & $0.03^{+0.02}_{-0.01}$            & $0.05^{+0.03}_{-0.01}$            & $0.05^{+0.02}_{-0.02}$        & $0.22^{+0.11}_{-0.07}$        & $0.9^{+0.5}_{-0.3}$           \\
                             & \multirow{3}{*}{No HG}   & $1.0$ TeV            & $90^{+50}_{-30}$                          & $0.020^{+0.010}_{-0.007}$         & $0.027^{+0.014}_{-0.009}$         & $0.04^{+0.02}_{-0.01}$            & $0.04^{+0.03}_{-0.01}$        & $0.19^{+0.09}_{-0.07}$        & $0.7^{+0.4}_{-0.2}$           \\
                             &                          & $1.2$ TeV            & $40^{+20}_{-10}$                          & $0.022^{+0.011}_{-0.007}$         & $0.03^{+0.02}_{-0.01}$            & $0.05^{+0.02}_{-0.02}$            & $0.05^{+0.02}_{-0.02}$        & $0.22^{+0.11}_{-0.07}$        & $0.9^{+0.4}_{-0.3}$           \\
                             &                          & $1.4$ TeV            & $50^{+20}_{-20}$                          & $0.023^{+0.012}_{-0.008}$         & $0.03^{+0.02}_{-0.01}$            & $0.05^{+0.03}_{-0.02}$            & $0.05^{+0.02}_{-0.02}$        & $0.22^{+0.11}_{-0.08}$        & $0.9^{+0.5}_{-0.3}$           \\ 
\hline
\multicolumn{3}{c}{Fit to $H_{T}$ distribution} & $90^{+40}_{-20}$ & $0.11^{+0.04}_{-0.04}$ & $0.11^{+0.05}_{-0.03}$ & $0.12^{+0.05}_{-0.04}$ & $0.3^{+0.1}_{-0.1}$ & $0.8^{+0.3}_{-0.2}$ & $1.7^{+0.7}_{-0.5}$ \\
    \end{tabular}
  \end{ruledtabular}
\end{table}

\begin{table}[h]
  \caption{\label{tab:limits_normalised}Normalised limits obtained for all combinations of training and testing signals.}
  \begin{ruledtabular}
    \begin{tabular}{ccclllllll}
      \multicolumn{1}{l}{}   & \multicolumn{1}{l}{}     & \multicolumn{1}{l}{} & \multicolumn{7}{c}{Test}                                                                                                                                                                                                                              \\
      \multicolumn{1}{l}{}   & \multicolumn{1}{l}{}     & \multicolumn{1}{l}{} & \multicolumn{1}{c}{\multirow{2}{*}{FCNC}} & \multicolumn{3}{c}{HG}            & \multicolumn{3}{c}{No HG}                                                                                                                                             \\
      \multicolumn{1}{l}{}   & \multicolumn{1}{l}{}     & \multicolumn{1}{l}{} & \multicolumn{1}{c}{}                      & \multicolumn{1}{c}{HG, $1.0$ TeV} & \multicolumn{1}{c}{HG, $1.2$ TeV} & \multicolumn{1}{c}{HG, $1.4$ TeV} & \multicolumn{1}{c}{$1.0$ TeV} & \multicolumn{1}{c}{$1.2$ TeV} & \multicolumn{1}{c}{$1.4$ TeV} \\
      \hline
      \multirow{7}{*}{Train} & \multicolumn{2}{c}{FCNC} & $1.0^{+0.4}_{-0.3}$  & $5^{+2}_{-2}$                             & $6^{+2}_{-2}$                     & $4^{+2}_{-1}$                     & $9^{+4}_{-3}$                     & $6^{+2}_{-2}$                 & $4^{+2}_{-1}$                                                 \\
                             & \multirow{3}{*}{HG}      & $1.0$ TeV            & $9^{+4}_{-3}$                             & $1.0^{+0.5}_{-0.3}$               & $1.3^{+0.7}_{-0.4}$               & $1.2^{+0.6}_{-0.4}$               & $1.3^{+0.7}_{-0.4}$           & $1.2^{+0.6}_{-0.4}$           & $1.3^{+0.7}_{-0.4}$           \\
                             &                          & $1.2$ TeV            & $8^{+4}_{-2}$                             & $0.8^{+0.4}_{-0.2}$               & $1.0^{+0.5}_{-0.3}$               & $1.0^{+0.5}_{-0.3}$               & $1.1^{+0.5}_{-0.4}$           & $1.0^{+0.5}_{-0.3}$           & $1.0^{+0.5}_{-0.3}$           \\
                             &                          & $1.4$ TeV            & $7^{+3}_{-2}$                             & $0.8^{+0.4}_{-0.3}$               & $1.0^{+0.5}_{-0.3}$               & $1.0^{+0.5}_{-0.3}$               & $1.1^{+0.6}_{-0.4}$           & $1.0^{+0.5}_{-0.3}$           & $1.0^{+0.5}_{-0.4}$           \\
                             & \multirow{3}{*}{No HG}   & $1.0$ TeV            & $20^{+9}_{-5}$                            & $0.7^{+0.4}_{-0.2}$               & $0.8^{+0.4}_{-0.3}$               & $0.8^{+0.4}_{-0.3}$               & $1.0^{+0.5}_{-0.3}$           & $0.9^{+0.4}_{-0.3}$           & $0.8^{+0.4}_{-0.3}$           \\
                             &                          & $1.2$ TeV            & $7^{+3}_{-2}$                             & $0.8^{+0.4}_{-0.2}$               & $1.0^{+0.5}_{-0.3}$               & $0.9^{+0.5}_{-0.3}$               & $1.1^{+0.5}_{-0.4}$           & $1.0^{+0.5}_{-0.3}$           & $1.0^{+0.5}_{-0.3}$           \\
                             &                          & $1.4$ TeV            & $9^{+4}_{-3}$                             & $0.8^{+0.4}_{-0.3}$               & $1.0^{+0.5}_{-0.3}$               & $1.0^{+0.5}_{-0.3}$               & $1.1^{+0.6}_{-0.3}$           & $1.0^{+0.5}_{-0.3}$           & $1.0^{+0.5}_{+0.3}$           \\
    \end{tabular}
  \end{ruledtabular}
\end{table}

\twocolumngrid

\begin{figure}[htb]
  \caption{\label{fig:limits_normalised_heatmap}Heatmap of the central value of the normalised limits obtained for all combinations of training and testing signals, as presented in \cref{tab:limits_normalised}.}
  \includegraphics[width=0.45\textwidth]{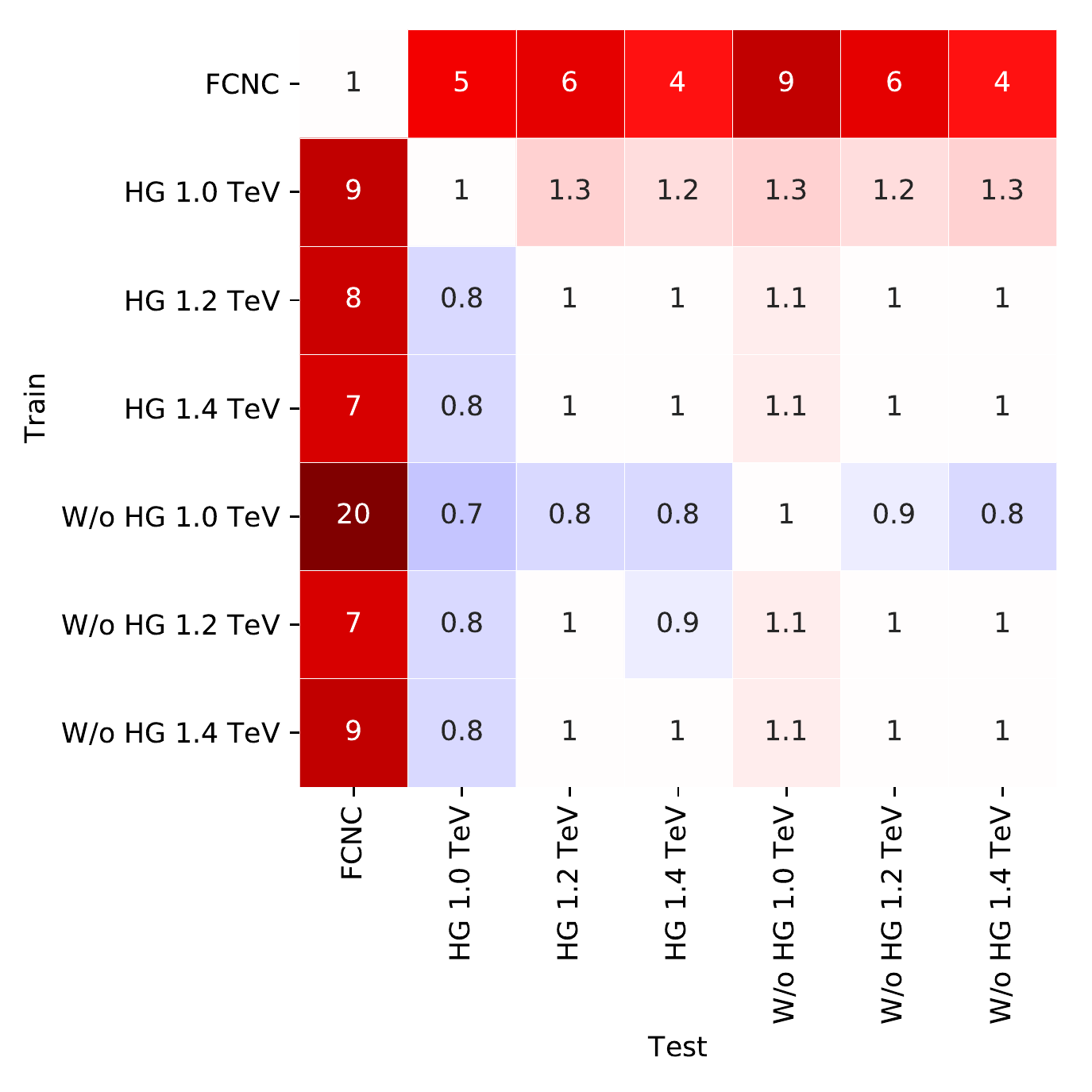}
\end{figure}

In order to better assess the impact on the limits as we use a network trained on a different signal, we normalise the columns with respect to the value of $\mu$ obtained by the network trained on the respective signal. These results are presented in \cref{tab:limits_normalised}. We also present the central values of this table as a heatmap in \cref{fig:limits_normalised_heatmap}. With the normalised results it becomes more clear that for the VLT signals the limits are insensitive within uncertainties to the specific signal used to train the network. Perhaps more surprisingly, the same holds when we use DNN trained on signals with heavy gluon on signals without and vice-versa, as signals produced from a heavy gluon will have different kinematics to those produced from a SM gluon.

Although there is a clear degradation of the upper limits of the FCNC signal when using a network trained on VLT signals, the change of the upper limits on those signals using the FCNC trained models is not as severe. This suggests that even in a limiting case where the signal is very similar to SM processes, such as the FCNC, it can still be used to train a network that can discriminate background from other much different signals. This might indicate that, for a DNN, signals are more alike to each other than to background, which can be interpreted as follows. During training a DNN will learn a decision function that defines a separation hypersurface between the regions supporting signal and background events in the high dimensional input space. While different signals are supported at different regions the background is common to all of them, which leads to each DNN trained on a specific signal to learn similar ways of isolating large parts of the background. It is this shared learned representational capacity of the background that is being transferred when we apply a trained DNN on a new signal.

\section{Conclusion}

In this work we set out to explore the transferability of DNN trained to discriminate between signal and background using reconstructed physical observables. Here we approached the notion of transferability as how a model trained on a given signal performs when discriminating between signal and background on a different signal. The different signals fell under three classes: a FCNC model without new heavy states, VLT states produced by SM gluon, and VLT states produced by a new heavy gluon. For the latter two cases we considered three different points of the parameter space where we varied the masses of the VLT states.

We have shown that the upper limits for the physical models with the VLT states are insensitive within uncertainties to which network discriminator we trained, as long as it was trained on a signal with VLT states. These results highlight that discriminative DNN are highly transferable across signals with similar new heavy states across different points of their parameter space. It should also be noted that the same transferability is observed between signals corresponding to significantly different models, where important kinematic differences between the reconstructed final state objects are expected.

Furthermore, we showed that networks trained in any signal can provide similar or greater sensitivity than performing limits on the fit to the $H_T$ distribution, demonstrating the aforementioned transferability of a trained DNN to new signals. More specifically, we showed that the FCNC signal, being the most different of the seven, can still be used to train a DNN that provides discriminative power to other signals that are kinematically very different. This suggests that in their training, DNNs learn features that are able to identify background, which is the same for all new physics. Ultimately, this means that the usage of DNNs in the discriminating step of an analysis might help to find new physics not assumed in its training. This fosters the idea that deep learning may contribute to a novel framework for generic searches, providing a powerful way to increase the sensitivity of searches for new physics phenomena at colliders. Future studies, beyond the scope of the present paper, can be done to verify the obtained results using detailed simulations of the LHC experiments. Also, it would be interesting to compare DNNs with other machine learning methods, also in terms of transferability of the results.

\section*{Acknowledgements}
We would like to thank A. Peixoto and J. Santiago for useful discussions and help with signal generation. We also acknowledge the support from FCT Portugal, Lisboa2020, Compete2020, Portugal2020 and FEDER
under the project PTDC/FIS-PAR/29147/2017 and through the grant PD/BD/135435/2017. The computational part of this work was supported by INCD (funded by FCT and FEDER under the project 01/SAICT/2016 nº 022153) and by the Minho Advanced Computing Center (MACC). The Titan Xp GPU card used for the training of the Deep Neural Networks developed for this project was kindly donated by the NVIDIA Corporation.

\section*{References}
\bibliography{notes}{}
\bibliographystyle{unsrt}

\end{document}